\documentclass[letterpaper,english,reprint, aps]{revtex4-1}
\usepackage[T1]{fontenc}
\usepackage[latin9]{inputenc}
\usepackage{color}
\usepackage{mathtools}
\usepackage{amsmath}
\usepackage{amssymb}
\usepackage{graphicx}

\makeatletter

\pdfpageheight\paperheight
\pdfpagewidth\paperwidth

\makeatother

\usepackage{babel}
\begin{document}

\preprint{APS/123-QED}

\title{Photoelectric converters with quantum coherence}

\author{Shan-He Su}

\affiliation{Beijing Computational Science Research Center, Beijing 100084, People's
Republic of China.}

\author{Sheng-Wen Li}

\affiliation{Institute of Quantum Science and Engineering, Texas A$\&$M University,
College Station, Texas 77843, USA.}

\author{Jin-Can Chen}

\affiliation{Department of Physics, Xiamen University, Xiamen 361005, People's
Republic of China.}

\author{Chang-Pu Sun}

\email{cpsun@csrc.ac.cn}

\affiliation{Beijing Computational Science Research Center, Beijing 100084, People's
Republic of China.}

\date{\today}
\begin{abstract}
Photon impingement is capable of liberating electrons in electronic
devices and driving the electron flux from the lower chemical potential
to higher chemical potential. Previous studies hinted that the thermodynamic
efficiency of a nano-sized photoelectric converter at maximum power
is bounded by the Curzon-Ahlborn efficiency $\eta_{CA}$. In this
study, we apply quantum effects to design a photoelectric converter
based on a three-level quantum dot (QD) interacting with fermionic
baths and photons. We show that, by adopting a pair of suitable degenerate
states, quantum coherences induced by the couplings of quantum dots
(QDs) to sunlight and fermion baths can coexist steadily in nano-electronic
systems. Our analysis indicates that the efficiency at maximum power
is no more limited to $\eta_{CA}$ through manipulation of carefully
controlled quantum coherences. 
\end{abstract}
\maketitle

\section{INTRODUCTION}

Carnot's theorem states that all real heat engines operating between
two heat baths undergo irreversible processes and are less efficient
than a reversible heat engine, regardless of the working substance
used or the operation details. Numerous studies have attempted to
design more efficient heat engines and improve the work extraction
when quantum effects come into play \citep{key-1,key-2,key-3,key-4}.
Most of quantum thermodynamic studies only emphasized on achieving
a conversion efficiency limit, which is inevitably accompanied by
vanishing power output \citep{key-5,key-6}. More extensive research
needs to be conducted regarding the interdependence of efficiency
and power for practical applications. Based on the Newton heat transfer
law, Curzon and Ahlborn found that the efficiency at maximum power
of an endoreversible Carnot heat engine with irreversible heat transfer
processes is given by $\eta_{CA}=1-\sqrt{T_{c}/T_{h}}$ , where $T_{h}$
is the temperature of the heat source and $T_{c}$ is the temperature
of the heat sink \citep{key-7}. Other various thermodynamic machines
indicate that $\eta_{CA}$ gives a good approximation for estimating
the efficiency at maximum power \citep{key-8,key-9,key-10}. In particular,
Rutten et al. proved that the efficiency at maximum power of a nanosized
photoelectric converter can be well predicted by the Curzon and Ahlborn
efficiency \citep{key-11}. Only in the case of the strong coupling
condition between electron and heat flows and negligible nonradiative
effects, can the efficiency more closely approach to $\eta_{CA}$.

An interesting question arises here: might quantum coherence survive
stably in nano-electronic systems and help to increase the efficiency
at maximum power beyond the bound of the Curzon and Ahlborn efficiency?
By considering a three-level quantum dot (QD) in thermal contact with
two boson reservoirs, Li et al. confirmed that the interferences of
two transitions in a non-equilibrium environment can give rise to
non-vanishing steady quantum coherence \citep{key-12}. Noise-induced
coherence is capable of breaking the detailed balance condition and
enhancing the laser power of a quantum heat engine \citep{key-13,key-14}.
The efficiency at maximum power of the laser quantum heat engine has
been shown to depend on the proper adjustment of the coherence parameters
\citep{key-15,key-16}. In these previous studies, the interaction
between quantum systems and bosonic baths plays a key role in generating
coherence. However, whether an electronic system in a fermionic environment
enables the realizations of steady coherence and performance improvement
is rarely discussed.

In this paper, in order to show that coherent transitions induced
by the couplings of quantum dots (QDs) to sunlight and fermion baths
can coexist to promote the potential of light harvesting, we propose
a experimentally feasible model of a nano-photoelectric converter.
We will focus on the condition to effectively increase the efficiency
at maximum power beyond the bound of Curzon-Ahlborn efficiency. The
contents are organized as follows: In Section II, the general model
of the converter is briefly described. In section III, the motion
equation of the QD is analytically computed. In section IV, the thermodynamic
quantities at steady state are derived. In Section V, the performance
characteristics of the photoelectric converter are revealed by numerical
calculation.

\begin{figure}
\includegraphics[scale=0.4]{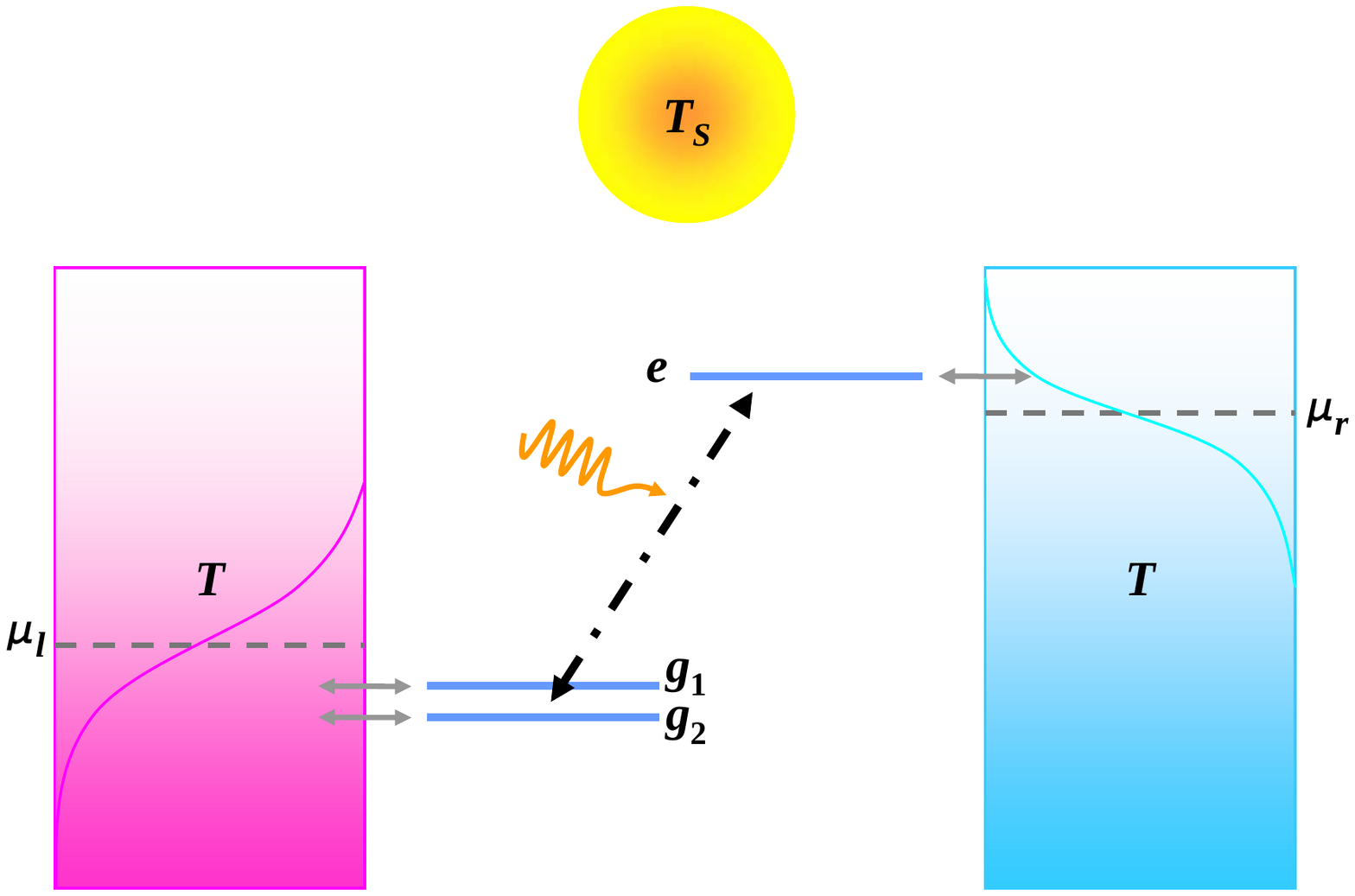}

\caption{Schematic of a photoelectric converter composed of a three-level QD.
The degenerate ground states $|g_{1}\rangle$ and $|g_{2}\rangle$
are coupled to the left-fermionic bath, while the excited state $|e\rangle$
is coupled to the right-fermionic bath. The two fermionic baths are
maintained at the same temperature $T$ but have different chemical
potentials $\mu_{l}$ and $\mu_{r}=\mu_{l}+qV$ due to the applied
voltage ($q$ is the elementary charge). Transitions between the ground
states and the excited state are induced by photons with temperature
$T_{S}$ (curved arrow).}
\end{figure}

\section{GENERAL DESCRIPTION OF THE MODEL}

The system schematic (Figure 1) of a photoelectric converter consists
of a three-level quantum dot (QD) contacted with two fermionic baths
and photons. The three-level QD is modeled by the Hamiltonian 

\begin{equation}
H_{S}=\sum_{i=g_{1},g_{2},e,0}\varepsilon_{i}|i\rangle\langle i|,
\end{equation}
with $|0\rangle$ being the state for no electron in the QD. $|g_{1}\rangle$,
$|g_{2}\rangle$, and $|e\rangle$ represent one-electron states in
levels $\varepsilon_{g_{1}}$, $\varepsilon_{g_{2}}$, or $\varepsilon_{e}$,
respectively. We assume that Coulomb repulsions prevent two electrons
to be simultaneously present in the QD \citep{key-17}. One electron
is firstly transferred from the left-fermionic bath to the ground
state $|g_{1}\rangle$ or $|g_{2}\rangle$ via the QD-bath coupling,
and then has a probability to be pumped to the excited state $|e\rangle$
due to the incoming solar radiation. The excited state $|e\rangle$
is coupled to the fermionic bath characterized by temperature $T$
and chemical potential $\mu_{r}$. 

The Hamiltonian of the sunlight radiation ($R{}_{P}$) is given by

\begin{equation}
H_{P}=\sum_{k\text{\ensuremath{\in R_{P}}}}\omega_{k}a_{k}^{\dagger}a_{k},
\end{equation}
where $\omega_{k}$ is the eigenfrequency of the radiated electromagnetic
wave described by the creation (annihilation) operator $a{}_{k}^{\dagger}$
$(a{}_{k})$. Similarly, the Hamiltonians of the fermionic baths ($R_{l}$
and $R_{r}$) are given by

\begin{equation}
H_{F}^{l,r}=\sum_{v\in R_{l,r}}\omega_{v}c_{v}^{\dagger}c_{v}.
\end{equation}
Here, $c{}_{v}^{\dagger}$ $(c{}_{v})$ is the electron creation (annihilation)
operator of the mode $\omega_{v}$ in $R_{l}$ or $R_{r}$. The two
ferminoic baths stand for the n-and p-type semiconductor electrodes
of the photoelectric converter.

The interaction between the QD and the environment reads $H_{I}=H_{I}^{l}+H_{I}^{r}+H_{I}^{P}$
with each term defined by

\begin{equation}
H_{I}^{l}=\underset{i=1,2}{\sum}\underset{v\in R_{l}}{\sum}(T_{v_{i}}c_{v}|g_{i}\rangle\langle0|+h.c.),
\end{equation}

\begin{equation}
H_{I}^{r}=\underset{v\in R_{r}}{\sum}(T_{v_{e}}c_{v}|e\rangle\langle0|+h.c.),
\end{equation}
and
\begin{equation}
H_{I}^{P}=\underset{i=1,2}{\sum}\underset{k\in R_{P}}{\sum}(g_{k_{i}}a_{k}|e\rangle\langle g_{i}|+h.c.),
\end{equation}
where $T_{v_{i}}$, $T_{v_{e}}$, and $g_{k_{i}}$ denote the coupling
strength of the transitions between the QD and the left-fermionic
bath, the right-fermionic bath and photons, respectively.

\section{MOTION EQUANTION OF THE QUANTUM DOT}

The three-level QD can be viewed as an open quantum system vulnerable
to interactions with the environment. Making the Born-Markov approximation,
which involves assuming that the environment is time-independent and
the environment correlations decay rapidly in comparison to the typical
timescale of the system evolution \citep{key-18}, we derive the equation
of motion for the density operator $\rho$ in a Lindblad-like form

\begin{equation}
\overset{.}{\rho}=i[\rho,H_{S}]+\mathcal{L_{\mathrm{\mathit{P}}}\mathrm{[\rho]}}+\mathcal{L_{\mathrm{\mathit{l}}}\mathrm{[\rho]}}+\mathcal{L_{\mathrm{\mathit{r}}}\mathrm{[\rho]}\mathrm{.}}
\end{equation}
 The dissipative part in the master equation can be generalized into
three individual elements including the dampings through the photons
and the two fermionic baths. For the photon excitation, the dissipation
operator $\mathit{\mathcal{L}_{P}}[\rho]$ depends on the Bose-Einstein
statistics of the photons and is given by

\begin{align}
\mathit{\mathcal{L}_{P}}[\rho] & =i[\rho,H_{CP}]\nonumber \\
 & +\underset{i,j=1}{\overset{2}{\sum}\lbrace[}B_{ij}^{+}(\varepsilon_{j})+B_{ij}^{+}(\varepsilon_{i})][\sigma_{Pi}^{\dagger}\rho\sigma_{Pj}-\frac{1}{2}\{\rho,\sigma_{Pj}\sigma_{Pi}^{\dagger}\}_{+}]\nonumber \\
 & +[{\textstyle B_{ji}^{-}(\varepsilon_{j})+B_{ji}^{-}(\varepsilon_{i})][\sigma_{Pi}\rho\sigma_{Pj}^{\dagger}-\frac{1}{2}\{\rho,\sigma_{Pj}^{\dagger}\sigma_{Pi}\}_{+}]}\rbrace,
\end{align}
where $\sigma_{Pi}=|g_{i}\rangle\langle e|$, $B_{ij}^{+}(\varepsilon_{j})=\gamma_{ij}^{P}(\varepsilon_{j})n(x_{j})$
and $B_{ij}^{-}(\varepsilon_{j})=\gamma_{ij}^{P}(\varepsilon_{j})[1+n(x_{j})]$
are the dissipation rates with $\gamma_{ij}^{P}(\omega)=\pi\sum g_{k_{i}}g_{k_{j}}^{\ast}\delta(\varepsilon-\omega)=[\gamma_{ji}^{P}(\omega)]^{\ast}$;
$n(x)=[exp(x)-1]^{-1}$ is the Bose-Einstein distribution with scaled
energy $x_{j}=\varepsilon_{j}/(k_{B}T_{P})$ and $k_{B}$ is the Boltzmann
constant. The energy difference of each transition is defined as $\varepsilon_{j}=\varepsilon_{e}-\varepsilon_{g_{j}}$.
The left-fermionic bath is coupled to the ground states $|g_{1}\rangle$
and $|g_{2}\rangle$. The corresponding dissipation operator is then
expressed as

\begin{align}
\mathit{\mathcal{L}_{l}}[\rho] & =i[\rho,H_{Cl}]\nonumber \\
 & +\underset{i,j=1}{\overset{2}{\sum}\lbrace[}F_{ij}^{l+}(\varepsilon_{g_{j}})+F_{ij}^{l+}(\varepsilon_{g_{i}})][\sigma_{li}^{\dagger}\rho\sigma_{lj}-\frac{1}{2}\{\rho,\sigma_{lj}\sigma_{li}^{\dagger}\}_{+}]\nonumber \\
 & +[{\textstyle F_{ji}^{l-}(\varepsilon_{g_{j}})+F_{ji}^{l-}(\varepsilon_{g_{i}})][\sigma_{li}\rho\sigma_{lj}^{\dagger}-\frac{1}{2}\{\rho,\sigma_{lj}^{\dagger}\sigma_{li}\}_{+}]}\rbrace.
\end{align}
where $\sigma_{li}=|0\rangle\langle g_{i}|$, $F_{ij}^{l+}(\varepsilon_{g_{j}})=\gamma_{ij}^{l}(\varepsilon_{g_{j}})f(x_{g_{j}})$
and $F_{ij}^{l-}(\varepsilon_{g_{j}})=\gamma_{ij}^{l}(\varepsilon_{g_{j}})[1-f(x_{g_{j}})]$
with $\gamma_{ij}^{l}(\omega)=\pi\sum T_{v_{i}}T_{v_{j}}^{\ast}\delta(\varepsilon-\omega)=[\gamma_{ji}^{l}(\omega)]^{\ast}$;
$f(x)=[exp(x)+1]^{-1}$ is the Fermi distribution with scaled energies
of the ground states $x_{g_{j}}=(\varepsilon_{g_{j}}-\mu_{l})/(k_{B}T)$.
The dissipation operator describes the coupling between the right-fermionic
bath and the excited state is

\begin{align}
\mathit{\mathcal{L}_{r}}[\rho]=F^{r+}(\varepsilon_{e})[2\sigma_{re}^{\dagger}\rho\sigma_{re}-\sigma_{re}\sigma_{re}^{\dagger}\rho-\rho\sigma_{re}\sigma_{re}^{\dagger}]\nonumber \\
+F^{r-}(\varepsilon_{e})[2\sigma_{re}\rho\sigma_{re}^{\dagger}-\sigma_{re}^{\dagger}\sigma_{re}\rho-\rho\sigma_{re}^{\dagger}\sigma_{re}]
\end{align}
where $\sigma_{re}=|0\rangle\langle e|$. $F^{r+}(\varepsilon_{e})=\gamma^{r}(\varepsilon_{e})f(x_{r})$
and $F^{r-}=\gamma^{r}(\varepsilon_{e})[1-f(x_{r})]$ with $\gamma^{r}(\omega)=\pi\sum T_{v_{e}}T_{v_{e}}^{\ast}\delta(\varepsilon-\omega)=[\gamma^{r}(\omega)]^{\ast}$.
$x_{r}=(\varepsilon_{e}-\mu_{r})/(k_{B}T)$ is the scaled energy of
the excited state.

Notice that the interference of coherent transitions can be simultaneously
induced by the photons and the left-fermionic bath, leading to two
different non-diagonal couplings given by

\begin{align}
H_{CP} & =\frac{1}{2i}\underset{i,j=1}{\overset{2}{\sum}\lbrace[}B_{ij}^{+}(\varepsilon_{i})-B_{ij}^{+}(\varepsilon_{j})]\sigma_{Pj}\sigma_{Pi}^{\dagger}\nonumber \\
 & +[{\textstyle B_{ji}^{-}(\varepsilon_{i})-B_{ji}^{-}(\varepsilon_{j})]\sigma_{Pj}^{\dagger}\sigma_{Pi}}\rbrace
\end{align}
and
\begin{align}
H_{Cl} & =\frac{1}{2i}\underset{i,j=1}{\overset{2}{\sum}\lbrace}[F_{ij}^{l+}(\varepsilon_{g_{i}})-F_{ij}^{l+}(\varepsilon_{g_{j}})]\sigma_{lj}\sigma_{li}^{\dagger}\nonumber \\
 & +[{\textstyle F_{ji}^{l-}(\varepsilon_{g_{i}})-F_{ji}^{l-}(\varepsilon_{g_{j}})]\sigma_{lj}^{\dagger}\sigma_{li}\rbrace}.
\end{align}

According to Eqs. (7)-(10), we have a coupled set of equations describing
the dynamics of the populations, $\rho_{i}=\langle g_{i}|\rho|g_{i}\rangle$,
$\rho_{e}=\langle e|\rho|e\rangle$, $\rho_{0}=\langle0|\rho|0\rangle$,
and the coherence, $\rho_{ij}=\langle g_{i}|\rho|g_{j}\rangle$, as
follows,

\begin{align}
\overset{.}{\rho}_{1} & =-2[B_{11}^{+}(\varepsilon_{1})+F_{11}^{l-}(\varepsilon_{g_{1}})]\rho_{1}+2B_{11}^{-}(\varepsilon_{1})\rho_{e}+2F_{11}^{l+}(\varepsilon_{g_{1}})\rho_{0}\nonumber \\
 & -[B_{12}^{+}(\varepsilon_{2})+F_{21}^{l-}(\varepsilon_{g_{2}})]\rho_{12}-[B_{21}^{+}(\varepsilon_{2})+F_{12}^{l-}(\varepsilon_{g_{2}})]\rho_{21}
\end{align}

\begin{align}
\overset{.}{\rho}_{2} & =-2[B_{22}^{+}(\varepsilon_{2})+F_{22}^{l-}(\varepsilon_{g_{2}})]\rho_{2}+2B_{22}^{-}(\varepsilon_{2})\rho_{e}+2F_{22}^{l+}(\varepsilon_{g_{2}})\rho_{0}\nonumber \\
 & -[B_{12}^{+}(\varepsilon_{1})+F_{21}^{l-}(\varepsilon_{g_{1}})]\rho_{12}-[B_{21}^{+}(\varepsilon_{1})+F_{12}^{l-}(\varepsilon_{g_{1}})]\rho_{21}
\end{align}

\begin{align}
\overset{.}{\rho}_{e} & =2B_{11}^{+}(\varepsilon_{1})\rho_{1}+2B_{22}^{+}(\varepsilon_{2})\rho_{2}-2[B_{11}^{-}(\varepsilon_{1})+B_{22}^{-}(\varepsilon_{2})\nonumber \\
 & +F^{r-}(\varepsilon_{e})]\rho_{e}+2F^{r+}(\varepsilon_{e})\rho_{0}+[B_{12}^{+}(\varepsilon_{1})+B_{12}^{+}(\varepsilon_{2})]\rho_{12}\nonumber \\
 & +[B_{21}^{+}(\varepsilon_{1})+B_{21}^{+}(\varepsilon_{2})]\rho_{21}
\end{align}

\begin{align}
\overset{.}{\rho}_{0} & =2F_{11}^{l-}(\varepsilon_{g_{1}})\rho_{1}+2F_{22}^{l-}(\varepsilon_{g_{2}})\rho_{2}+2F^{r-}(\varepsilon_{e})\rho_{e}-2[F_{11}^{l+}(\varepsilon_{g_{1}})\nonumber \\
 & +F_{22}^{l+}(\varepsilon_{g_{2}})+F^{r+}(\varepsilon_{e})]\rho_{0}+[F_{21}^{l-}(\varepsilon_{g_{1}})+F_{21}^{l-}(\varepsilon_{g_{2}})]\rho_{12}\nonumber \\
 & +[F_{12}^{l-}(\varepsilon_{g_{1}})+F_{12}^{l-}(\varepsilon_{g_{2}})]\rho_{21}
\end{align}

and
\begin{align}
\overset{.}{\rho}_{12} & =-[B_{21}^{+}(\varepsilon_{1})+F_{12}^{l-}(\varepsilon_{g_{1}})]\rho_{1}-[B_{21}^{+}(\varepsilon_{2})+F_{12}^{l-}(\varepsilon_{g_{2}})]\rho_{2}\nonumber \\
 & +[B_{21}^{-}(\varepsilon_{1})+B_{21}^{-}(\varepsilon_{2})]\rho_{e}+[F_{12}^{l+}(\varepsilon_{g_{1}})+F_{12}^{l+}(\varepsilon_{g_{2}})]\rho_{0}\nonumber \\
 & -[B_{11}^{+}(\varepsilon_{1})+B_{22}^{+}(\varepsilon_{2})+F_{11}^{l-}(\varepsilon_{g_{1}})+F_{22}^{l-}(\varepsilon_{g_{2}})+\tau]\rho_{12}\nonumber \\
 & +i\varDelta_{21}\rho_{12}.
\end{align}
Here, $\varDelta_{21}=\varepsilon_{g_{2}}-\varepsilon_{g_{1}}$ is
the energy difference of the two lower states $|g_{1}\rangle$ and
$|g_{2}\rangle$, and $\tau$ is phenomenologically introduced to
describe the decoherence rate due to the environment effects. The
equations for off-diagonal terms, e.g. $\langle g_{i}|\rho|e\rangle$
and $\langle e|\rho|0\rangle$, have been omitted except $\rho_{12}$,
since those terms only give the decay processes and do not affect
the steady-state solution. It is shown that the time evolutions of
the populations $\rho_{i}$, $\rho_{e}$, and $\rho_{0}$ are not
decoupled from that of the off-diagonal elements $\rho_{12}/\rho_{21}$.
The coherence $\rho_{12}/\rho_{21}$ may not vanish even in the steady
state after long time evolution. Specifically, we find that both QD-photon
coupling and QD-fermion coupling contribute to the coherent transitions.

\section{THERMODYNAMIC QUANTITIES AT STEADY STATE}

For degenerate lower levels $\varepsilon_{g_{1}}=\varepsilon_{g_{2}}=\varepsilon_{l}$
and symmetric couplings, we write the rates of transitions $|g_{1}\rangle\leftrightarrow|e\rangle$
and $|g_{2}\rangle\leftrightarrow|e\rangle$ as $\gamma_{11}^{P}(\varepsilon_{1})=\gamma_{22}^{P}(\varepsilon_{2})=\gamma^{P}$
and that of transitions $|g_{1}\rangle\leftrightarrow|0\rangle$ and
$|g_{2}\rangle\leftrightarrow|0\rangle$ as $\gamma_{11}^{l}(\varepsilon_{g_{1}})=\gamma_{22}^{l}(\varepsilon_{g_{2}})=\gamma^{l}$.
We also introduce two dimensionless parameters $r_{P}(=\gamma_{12}^{P}/\gamma^{P})$
and $r_{l}(=\gamma_{12}^{l}/\gamma^{l})$ to describe the strengths
of coherences, where superscripts $P$ and $l$ imply the coherent
transitions originating from the couplings to the photons and the
left-fermionic bath, respectively. Note that $0\leq r_{P},r_{l}$$\leq1$,
depending on the relative orientations of transition dipole vectors
\citep{key-15}. Setting $\stackrel{.}{\rho}=0$ and combining Eqs.
(13)-(17) with the conservative equation $\rho_{1}+\rho_{2}+\rho_{e}+\rho_{0}=1$,
the steady-state populations and coherence of the open quantum system
is obtained. The coherence is computed as

\begin{align}
\rho_{12} & =2\gamma^{P}(r_{P}-r_{l})\{n(x_{g})f(x_{l})-[1+n(x_{g})\nonumber \\
 & -f(x_{l})]f(x_{r})\}/\Omega,
\end{align}
where $\Omega$ is the normalization factor that ensures the sum of
probabilities to be equal to unity. Simplifying the numerator of Eq.
(18) to $1/2\gamma^{P}(r_{l}-r_{P})\textrm{Csch}(x_{g}/2)\textrm{Sech}(x_{l}/2)\textrm{Sech}(x_{r}/2)\textrm{Sinh}[(x_{g}+x_{l}-x_{r})/2)]$,
we identify that $\rho_{12}$ reduces to zero when $r_{P}=r_{l}$
and the quantum coherence will not affect the thermodynamics. This
phenomenon was observed in a four-level quantum heat engine for symmetric
coupling condition as well \citep{key-16}.

From the master equation, the changing rate of the electron number
in the three-level QD at time $t$ is

\begin{equation}
\overset{.}{\overline{N}}(t)=\mathrm{Tr}\{n\mathcal{L_{\mathrm{\mathit{l}}}\mathrm{[\rho]}}\}+\mathrm{Tr}\{n\mathcal{L_{\mathrm{\mathit{r}}}\mathrm{[\rho]}}\}\coloneqq J_{l}+J_{r}
\end{equation}
with the number operator $n=\sigma_{re}^{\dagger}\sigma_{re}+\sigma_{l1}^{\dagger}\sigma_{l1}+\sigma_{l2}^{\dagger}\sigma_{l2}$.
Thus, $J_{l}$ and $J_{r}$ are the currents exchanging with the left
and the right fermionic baths, which are given by

\begin{equation}
J_{l}=4\gamma^{l}f(x_{l})\rho_{0}-2\gamma^{l}[1-f(x_{l})]\{\rho_{1}+\rho_{2}+r_{l}\mathrm{Re[}\rho_{12}]\}
\end{equation}
and

\begin{equation}
J_{r}=2\gamma^{r}[1-f(x_{r})]\rho_{e}-2\gamma^{r}f(x_{r})\rho_{0}.
\end{equation}
The parameter, $x_{l}=(\varepsilon_{l}-\mu_{l})/(k_{B}T)$, is the
scaled energy of the degenerate ground states. In the stationary state
$(t\rightarrow\text{\ensuremath{\infty}})$, $\overset{.}{N}(t)=0$
such that $J_{l}=-J_{r}$. Eq. (20) indicates that adjusting the electron
current via quantum coherences allows for improving the performance
of the converter. The steady state energy fluxes are determined by
energy change of the three-level QD, i.e. $\overset{.}{E}(\text{\ensuremath{\infty}})=\mathrm{Tr}\{H_{S}\overset{.}{\rho}(\text{\ensuremath{\infty}})\}=\underset{\alpha}{\sum}\mathrm{Tr}\{H_{S}\mathcal{L}_{\alpha}[\rho(\text{\ensuremath{\infty}})]\}$
($\alpha=P$, $l$, and $r$ ). Neglecting the nonradiative recombination
processes \citep{key-11,key-20}, the net heat flux coming from the
sunlight $\overset{.}{Q}_{P}=\mathrm{Tr}\{H_{S}\mathcal{L}_{P}[\rho(\text{\ensuremath{\infty}})]\}=\varepsilon_{g}J$,
where $\varepsilon_{g}=\varepsilon_{e}-\varepsilon_{l}$ can be regarded
as the bandgap energy. The power $P$ generated by the photoelectric
converter to move electrons from the left-fermionic bath to the right-fermionic
bath yields

\begin{equation}
P=(\mu_{r}-\mu_{l})J=k_{B}T_{P}[x_{g}-(1-\eta_{c})(x_{r}-x_{l})]J
\end{equation}
with $x_{g}=\varepsilon_{g}/(k_{B}T_{P})$. The symbol $\eta_{c}$
denotes the Carnot efficiency and equals $1-T/T_{P}$. The efficiency
satisfying this conversion is then given by 
\begin{equation}
\eta=\frac{P}{\overset{.}{Q}_{P}}=\frac{(\mu_{r}-\mu_{l})J}{\varepsilon_{g}J}=1-(1-\eta_{c})\frac{(x_{r}-x_{l})}{x_{g}}.
\end{equation}

\begin{figure}[b]
\includegraphics[width=4.6cm,height=4.2cm]{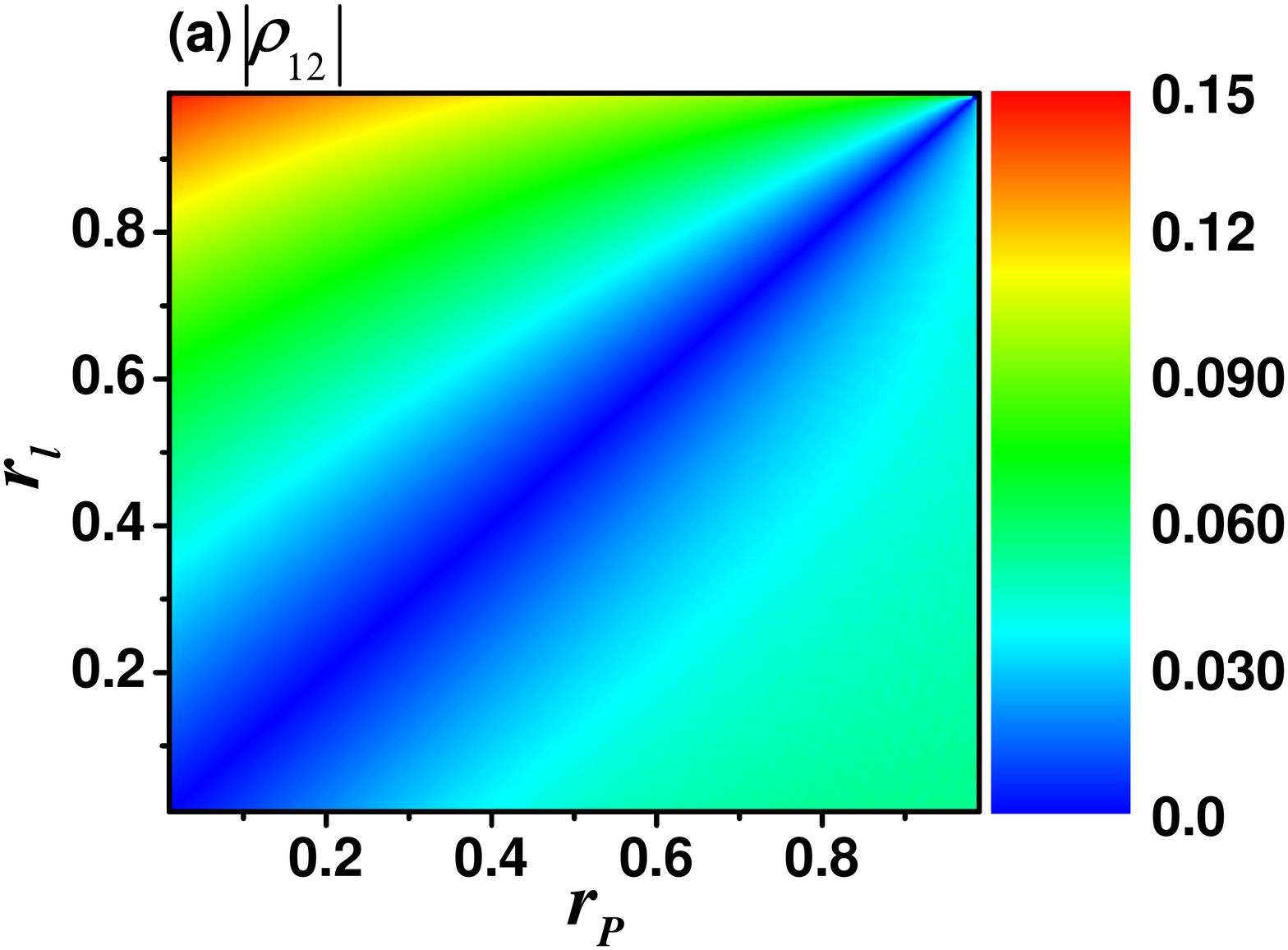}\includegraphics[width=4.6cm,height=4.2cm]{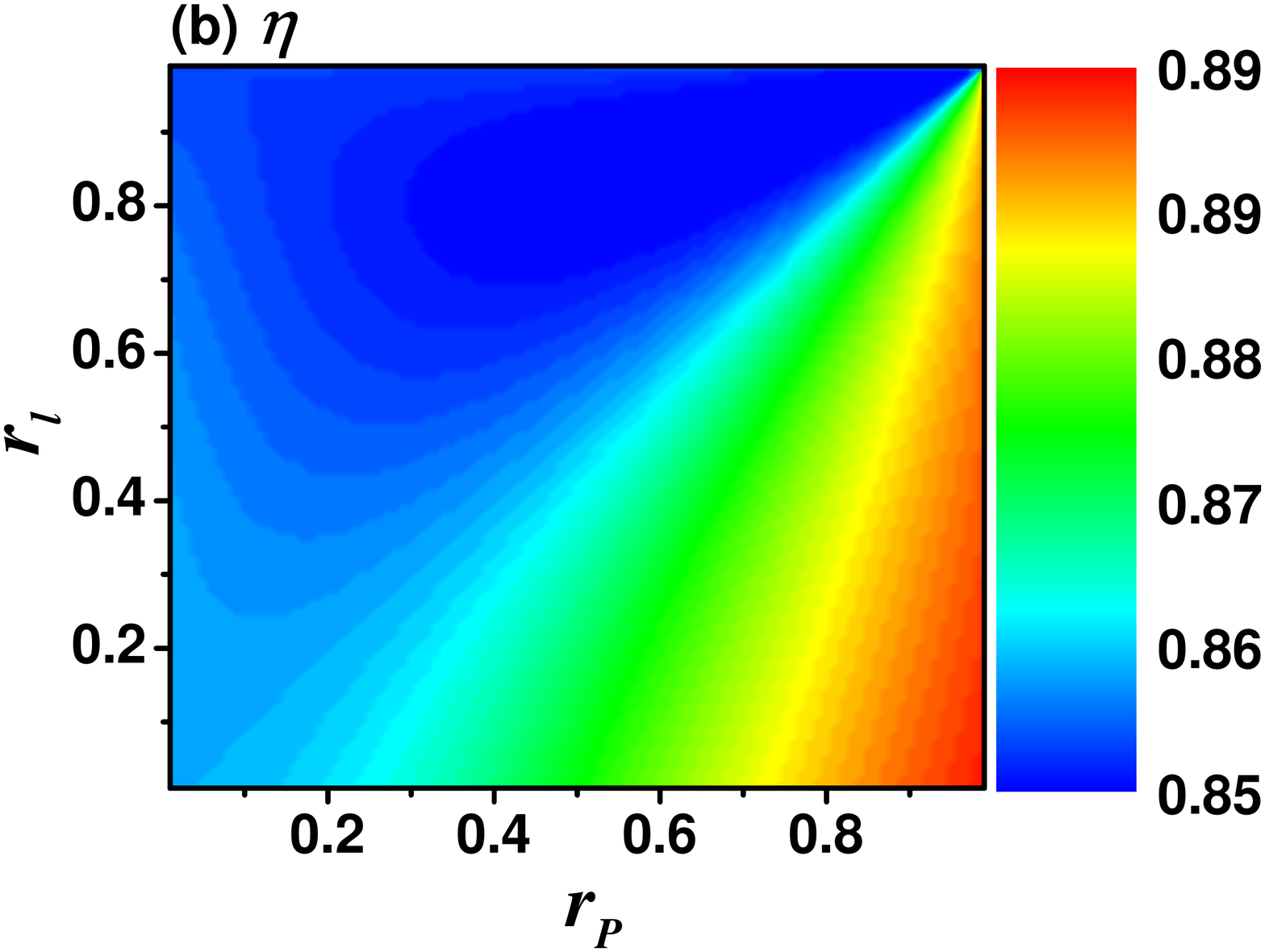}\caption{The absolute value of coherence (a) and the efficiency (b) as a function
of the dimensionless parameters $r_{P}$ and $r_{l}$, where $x_{g}=2,$
$\tau=0$, $T=295K$, and $T_{P}=5780K$. The optimal values of $x_{l}$
and $x_{r}$ have been computed numerically to maximize the power
output. }
\end{figure}

\begin{figure}[tb]

\includegraphics[width=9cm,height=6cm]{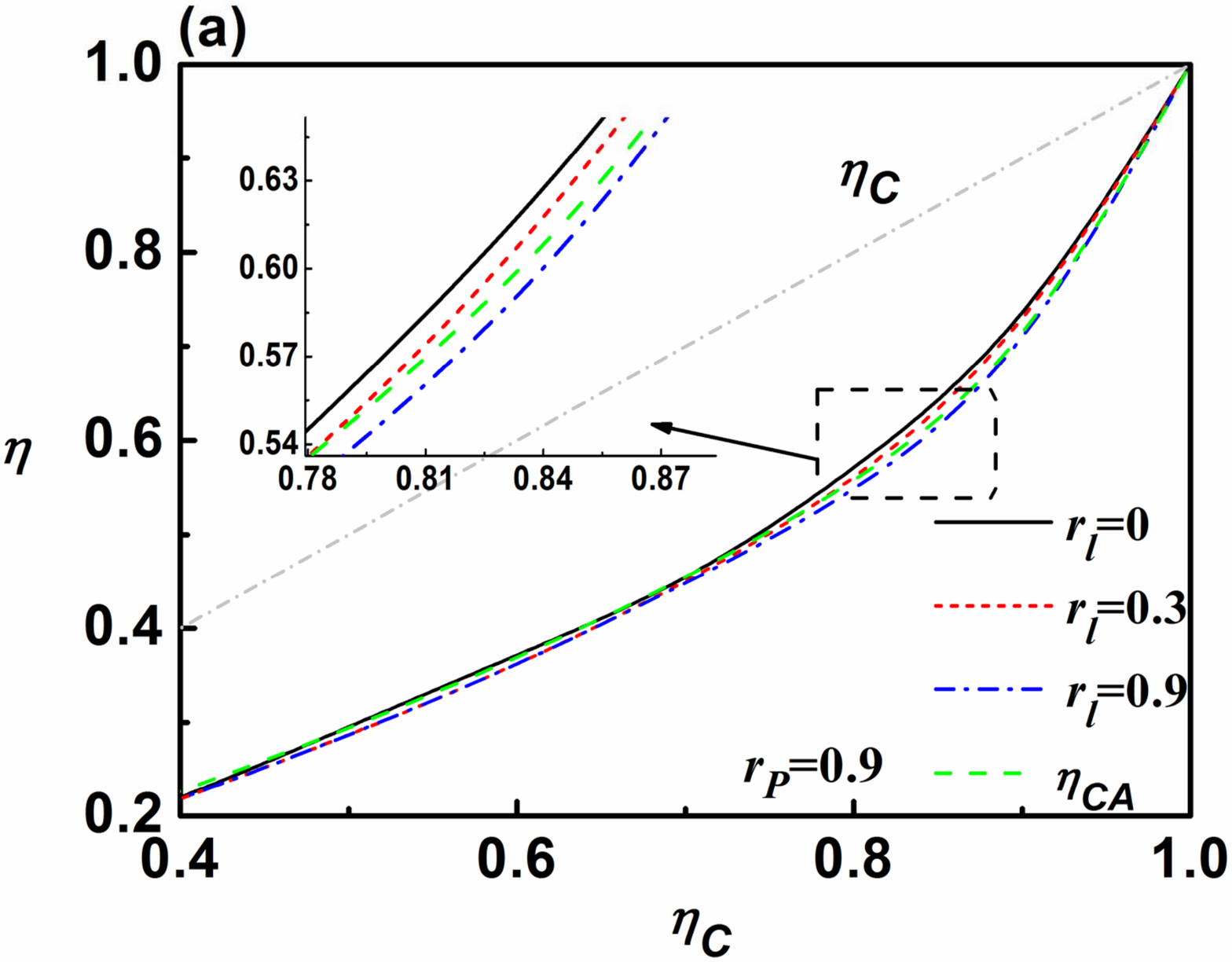}

\includegraphics[width=9cm,height=6cm]{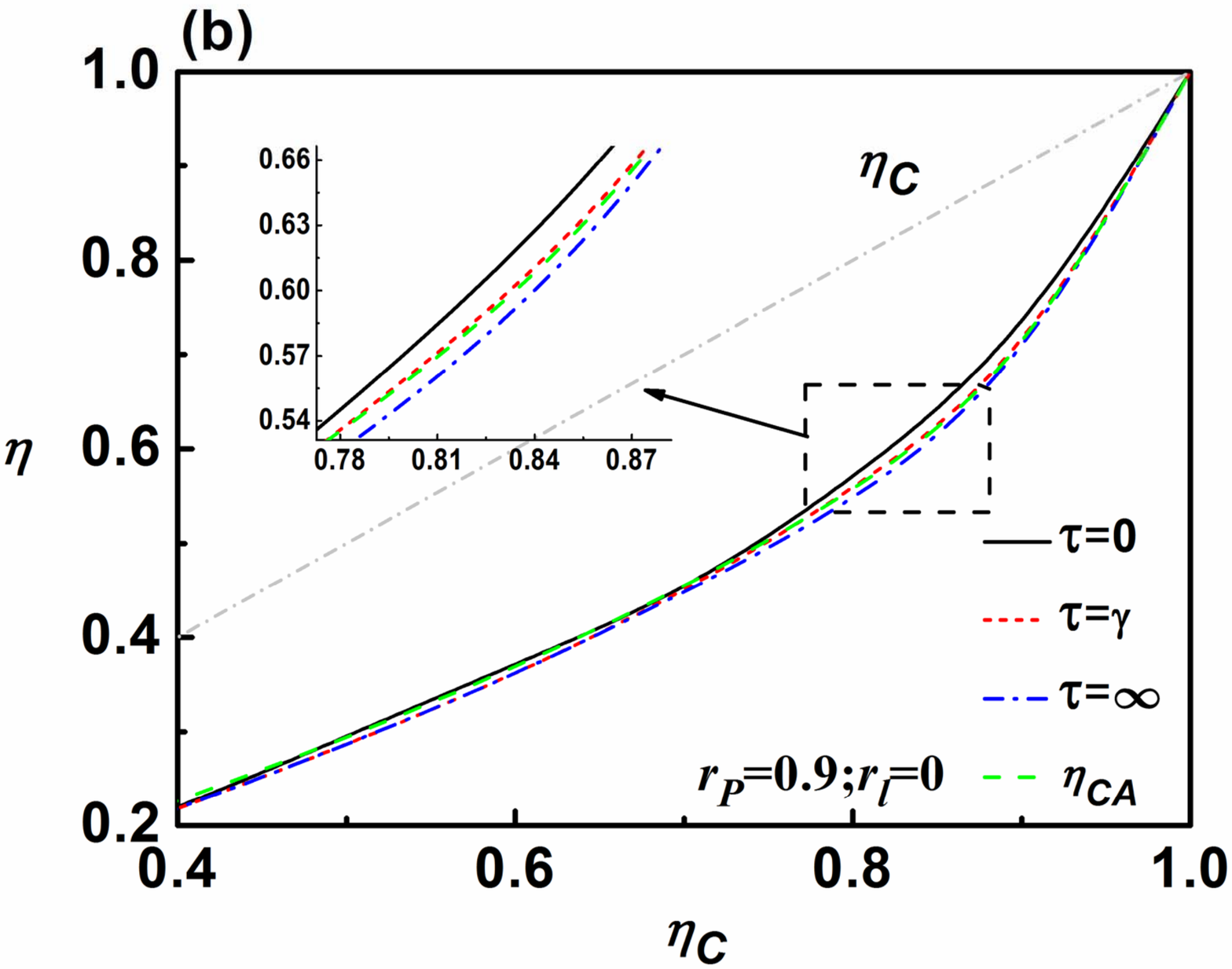}\caption{The efficiency at maximum power and the Curzon-Ahlborn efficiency
(green dashed line) as a function of Carnot efficiency $\eta_{c}$
for different values of $r_{l}$ (a) and $\tau$ (b). In Fig. 3a,
$r_{P}=0.9$ and $\tau=0$. In Fig. 3b, $r_{P}=0.9$ and $r_{l}=0$.
The inserted figure shows an enlargement of the representative part
of each plot.}
\end{figure}

\section{PERFORMANCE CHARACTERISTIC ANALYSIS}

In the following section, we address the question regarding the extent
to which the quantum nature of the converter affects the photoelectric
conversion efficiency, a topic which is beyond the reach of the model
presented in Refs. {[}11{]}. The formalism obtained here will allow
us to access how coherences can lead to an enhancement of the power
and the efficiency. To do so, we parameterize the transition rates
$\gamma^{P}=\gamma^{l}=\gamma^{r}$=$\gamma$ without loss of generality.

Figures 2a and 2b show the contour plots of the absolute value of
coherence $\mid\rho_{12}\mid$ and the efficiency $\eta$ versus $r_{P}$
and $r_{l}$, where the power has been optimized with respect to $x_{l}$
and $x_{r}$. In accordance with the requirements set out in the analytical
method (Eq. 18), Fig. 2a shows that the quantum coherence vanishes
if $r_{P}=r_{l}$, resulting in low efficiency less than $0.87$.
To enhance $\eta$ in the presence of coherence $(\mid\rho_{12}\mid\neq0)$,
$r_{P}$ and $r_{l}$ should be designed to be different from each
other. However, we notice that $\eta$ and $\mid\rho_{12}\mid$ may
not always increase or decrease together, which means that $\eta$
is not only restricted by the magnitude of coherence. Comparing Fig.
2a with Fig. 2b, we find that $\eta$ can be largely enhanced in the
range of $r_{P}>r_{l}$. This condition suggests that increasing the
coherence coupling between the QD and the photons and $\mid\rho_{12}\mid$
will concurrently benefit the performance of the photoelectric conversion.
There exists a perfect positive correlation between $\eta$ and $\mid\rho_{12}\mid$
when $r_{P}>r_{l}$ is satisfied.

Next, we maximize the power with respect to $x_{g}$ , $x_{l}$ ,
and $x_{r}$. In Figure 3a, the efficiency at maximum power is plotted
as a function of $\eta_{c}$ for given values of $r_{P}$. Figure
3a shows that the efficiency at maximum power increases with a decrease
in the parameter $r_{l}$, as expected. When $r_{P}=r_{l}=0.9$ (dash-dotted
line), the efficiency remains close to the Curzon-Ahlborn efficiency
for almost all values of $\eta_{c}$. Slight lower efficiencies are
observed only far from equilibrium where $\eta_{c}$ is large. These
features have also been addressed in other approaches to non-equilibrium
thermodynamics, such as Brownian heat engine \citep{key-21,key-22},
Feynman ratchet model \citep{key-23}, and thermoelectric device \citep{key-24},
and therefore point to a fundamental principle that can be associated
with the presence of inherently irreversible dynamics when device
is operated at maximum power. 

Intriguingly, we find that the efficiency at maximum power is not
limited by the Curzon-Ahlborn efficiency. For example, when $r_{l}=0.3$
(short-dashed line) or $r_{l}=0$ (solid line), the quantum coherence
appears and the efficiency at maximum power will exceed the bound
given by the Curzon-Ahlborn efficiency. These results are remarkable
in our model with quantum coherence. The quantum coherence will redistribute
the population in the three-level QD and accelerate the removal of
electrons, thus increasing the number of absorbed photons and reducing
recombination losses.

Finally, we consider the typical problem arising due to the decoherence.
Decoherence occurs when a system interacts with its environment in
a thermodynamically irreversible way. The decoherence processes can
drastically decrease an engine's efficiency. In Fig. 3b, the efficiency
at maximum power is plotted as a function of $\eta_{c}$ with $r_{P}=0.9$
and $r_{l}=0$. In the case that the decoherence rate is extremely
large, i.e., $\tau\rightarrow\infty$, the efficiency (dashed line)
again becomes slightly lower than the Curzon-Ahlborn efficiency. As
$\tau$ diminishes, we find that the efficiency increases monotonically.
The efficiency at maximum power is significantly higher than the Curzon-Ahlborn
efficiency when $\tau=0$.

\section{CONCLUDSIONS}

In summary, we propose a new type of photoelectric converter, which
consists of a three-level QD coupling to two fermionic baths and sunlight
radiation. It follows from the Born-Markov approximation that the
interference due to coherent transitions can be simultaneously induced
by the sunlight and the left-fermionic bath, leading to two different
non-diagonal Lamb shifts in the Lindblad-like master equation. The
results of the thermodynamic analysis show that the quantum coherence
is capable of improving the efficiency beyond the limit of a system
whose quantum effects are absent. The application of quantum mechanics
will bring new insight to understand the fundamantal problem in thermodynamics
when it is applied to the nano-electronic systems.
\begin{acknowledgments}
This work has been supported by the National Natural Science Foundation
of China (Grant Nos. 11421063 and 11534002), the National 973 program
(Grant Nos. 2012CB922104 and 2014CB921403), and the Postdoctoral Science
Foundation of China (Grant No. 2015M580964).\end{acknowledgments}


\begin{thebibliography}{99}
\bibitem{key-1}M. O. Scully, M. S. Zubairy, G. S. Agarwal, and H.
Walther, \textcolor{blue}{Science }\textbf{\textcolor{blue}{299,}}\textcolor{blue}{{}
862-864 (2003).}

\bibitem{key-2}J. Roßnagel, O. Abah, F. Schmidt-Kaler, K. Singer,
and E. Lutz, \textcolor{blue}{Phys. Rev. Lett. }\textbf{\textcolor{blue}{112,}}\textcolor{blue}{{}
030602 (2014).}

\bibitem{key-3}X. L. Huang, Tao Wang, and X. X. Yi, \textcolor{blue}{Phys.
Rev. E }\textbf{\textcolor{blue}{86,}}\textcolor{blue}{{} 051105 (2012).}

\bibitem{key-4}H. T. Quan, P. Zhang, and C. P. Sun, \textcolor{blue}{Phys.
Rev. E }\textbf{\textcolor{blue}{73,}}\textcolor{blue}{{} 036122 (2006).
}

\bibitem{key-5}R. S. Whitney, \textcolor{blue}{Phys. Rev. Lett. }\textbf{\textcolor{blue}{112,}}\textcolor{blue}{{}
130601 (2014).} 

\bibitem{key-6}M. Esposito, R. Kawai, K. Lindenberg, and C. Van den
Broeck, \textcolor{blue}{Phys. Rev. Lett. }\textbf{\textcolor{blue}{105,}}\textcolor{blue}{{}
150603 (2010)}. 

\bibitem{key-7}F.L. Curzon and B. Ahlborn, \textcolor{blue}{Am. J.
Phys. }\textbf{\textcolor{blue}{43,}}\textcolor{blue}{{} 22-24 (1975).}


\bibitem{key-8}J. Wang, Z. Ye, Y. Lai, W. Li, and J. He, \textcolor{blue}{Phys.
Rev. E }\textbf{\textcolor{blue}{91,}}\textcolor{blue}{{} 062134 (2015);}
 

\bibitem{key-9}F. Wu, J. He, Y. Ma, and J. Wang, \textcolor{blue}{Phys.
Rev. E }\textbf{\textcolor{blue}{90,}}\textcolor{blue}{{} 062134 (2014)}.


\bibitem{key-10}J. Guo, J. Wang, Y. Wang, and J. Chen, \textcolor{blue}{Phys.
Rev. E }\textbf{\textcolor{blue}{87,}}\textcolor{blue}{{} 012133 (2013).}


\bibitem{key-11}B. Rutten, M. Esposito, and B. Cleuren, \textcolor{blue}{Phys.
Rev. B }\textbf{\textcolor{blue}{80,}}\textcolor{blue}{{} 235122 (2009).}


\bibitem{key-12}S. W. Li, C. Y. Cai, and C. P. Sun, \textcolor{blue}{Ann.
Phys. -New York }\textbf{\textcolor{blue}{360,}}\textcolor{blue}{{}
19-32 (2015)}. 

\bibitem{key-13}M. O. Scully, K. R. Chapin, K. E. Dorfman, M. B.
Kim, and A. Svidzinsky, \textcolor{blue}{Proc. Natl. Acad. Sci. U.S.A.
}\textbf{\textcolor{blue}{108,}}\textcolor{blue}{{} 15097-15100 (2011).
}

\bibitem{key-14}K. E. Dorfman, D. V. Voronine, S. Mukamel, and M.
O. Scully, \textcolor{blue}{Proc. Natl. Acad. Sci. USA }\textbf{\textcolor{blue}{110,}}\textcolor{blue}{{}
2746-2751 (2013).}

\bibitem{key-15}U. Harbola, S. Rahav, and S. Mukamel, \textcolor{blue}{EPL
}\textbf{\textcolor{blue}{99,}}\textcolor{blue}{{} 50005 (2012). }

\bibitem{key-16}H. P. Goswami and U. Harbola, \textcolor{blue}{Phys.
Rev. A }\textbf{\textcolor{blue}{88,}}\textcolor{blue}{{} 013842 (2013).
}

\bibitem{key-17}B. Cleuren, B. Rutten, and C. Van den Broeck, \textcolor{blue}{Phys.
Rev. Lett. }\textbf{\textcolor{blue}{108,}}\textcolor{blue}{{} 120603
(2012).} 

\bibitem{key-18}H. P. Breuer and F. Petruccione, T\textcolor{black}{he
theory of open quantum systems} (Oxford University Press, Oxford,
2001).

\bibitem{key-20}J. Wang, Y. Lai, Z. Ye, J. He, Y. Ma, and Q. Liao,
\textcolor{blue}{Phys. Rev. E }\textbf{\textcolor{blue}{91,}}\textcolor{blue}{{}
050102(R) (2015).} 

\bibitem{key-21}V. Blickle and C. Bechinger, \textcolor{blue}{Nat.
Phys. }\textbf{\textcolor{blue}{8,}}\textcolor{blue}{{} 143\textendash 146
(2012).}  

\bibitem{key-22}Z. C. Tu, \textcolor{blue}{Phys. Rev. E }\textbf{\textcolor{blue}{89,}}\textcolor{blue}{{}
052148 (2014).} 

\bibitem{key-23}Z. C. Tu, \textcolor{blue}{J. Phys. A }\textbf{\textcolor{blue}{41,}}\textcolor{blue}{{}
312003 (2008). }

\bibitem{key-24}M Esposito, K Lindenberg, and C. Van den Broeck,
\textcolor{blue}{EPL }\textbf{\textcolor{blue}{85,}}\textcolor{blue}{{}
60010 (2009).} \end{thebibliography}
\end{document}